\documentstyle[preprint,aps]{revtex}

\def\hyp#1#2{${}^{#1}_{\Lambda}$#2}

\def\lnnn{$\Lambda$N$\to$NN }

\def\c12{$^{12}_\Lambda$C}
\def\b11{$^{11}_\Lambda$B}
\def\he5{$^{5}_\Lambda$He}

\def\dioh{$\Delta$I=1/2\ }
\def\dith{$\Delta$I=3/2\ }

\begin{document}
\draft
\title{
Violation of the $\Delta$I=1/2 rule in the nonmesonic weak decay \\
of hypernuclei}

\author{\underline {A. Parre\~no}\footnote{e-mail address:
assum@dirac.ecm.ub.es}, A. Ramos}
\address{
Departament d'Estructura i Constituents de la Mat\`eria, Universitat de
Barcelona,\\
Diagonal 647, 08028 Barcelona, Spain}
\author{C. Bennhold}
\address{
Center of Nuclear Studies, Department of Physics,
The George Washington University, Washington, DC 20052}
\author{K. Maltman}
\address{
Department of Mathematics and Statistics, York University, 4700 Keele
St., North York, Ontario, Canada M3J 1P3
}
\maketitle

\begin{abstract}
Violations of the $\Delta$I=1/2 rule are investigated in the nonmesonic
weak hypernuclear decay using a weak \lnnn transition potential
based on meson exchange.
While the weak $\Delta$I=3/2 couplings of
baryons to pseudoscalar mesons
are known to be very small,
the analogous couplings
of baryons to vector mesons
are not known experimentally.
These couplings have been evaluated
using the factorization approximation and
are found to produce potentially significant changes
in the predictions for
hypernuclear
decay observables in the case of ${}^{12}_\Lambda$C.
Within the uncertainties of the factorization approximation we find
that the proton-induced rate is affected by at most 10\%, while the
neutron-induced rate can change by up to a factor of two.
The asymmetry parameter is strongly affected as well.
\end{abstract}

\newpage

The observed dominance of strangeness-changing non-leptonic weak
interactions by $\Delta$I=1/2 transition operators in the two
processes studied to date (hyperon decay and $K\rightarrow 2\pi
,3\pi$) has often lead to the assumption that the \dioh rule will
turn out to be a universal feature of all such non-leptonic
interactions.  It is, however, not clear that this is necessarily
the case. Ref.~\onlinecite{MS94}, for example, from a quark model
perspective, suggests the possibility of significant \dith
contributions to transitions from the ${^1S}_0$ initial state in
$\Lambda$N$\rightarrow$NN. Similarly, relevant to meson exchange
model treatments of $\Lambda$N$\rightarrow$NN, it has been
shown\onlinecite{MS95} that there may be significant \dith
contributions to the weak vector meson-baryon couplings.  It is
important to stress, in this regard, that there is a qualitative
physical distinction between the weak baryon-pseudoscalar and weak
baryon-vector meson couplings. This follows from the fact that the
factorization contributions generated by those operators in the
QCD-enhanced effective weak Hamiltonian associated with penguin
graphs receive large ($\simeq 10$) enhancements in the case of the
weak baryon-pseudoscalar couplings (due to the different chiral
structure of the penguin-induced operators), in contrast to the
case, for example, of the weak $\Lambda$N$\rho$, $\Sigma$N$\rho$,
and $\Sigma$N$\omega$ couplings, for which the
corresponding factorization contributions actually
vanish\onlinecite{MS95}.  The observed \dioh dominance of the weak
baryon-pseudoscalar couplings should, thus, not be taken as
providing any evidence for a similar dominance of the baryon-vector
meson couplings by \dioh operators.  We will, in consequence, in
what follows, attempt to make estimates of the sizes of the weak
\dith contributions to the $\Lambda$N$\rho$, and
NNK$^*$ couplings relevant to $\Lambda$N$\rightarrow$NN, and
explore the influence of such terms on various hypernuclear decay
observables.

The starting point for our estimate of the \dith coupling
contributions is the factorization approximation.  As is well-
known, QCD corrections to the basic weak interactions produce an
effective weak Hamiltonian, ${\cal H}_{eff}$, which can be
evaluated using perturbative QCD, down to a scale $\sim 1\ $GeV.
The form of ${\cal H}_{eff}$ for the non-leptonic strangeness
changing weak interactions is then found to
be\onlinecite{heffref1,heffref2}:
\begin{equation}
{\cal H}_{eff}=-\sqrt{2}G_F\sin\theta_C\cos\theta_C
\sum^{6}_{i=1}c_i O_i
\label{QCD}
\end{equation}
where $G_F$ is the Fermi constant, $\theta_C$ the Cabbibo
angle and the operators $O_i$ have the form:
\begin{eqnarray}
&&O_1 = \bar{d}_L\gamma_{\mu}s_L\,\bar{u}_L\gamma^{\mu}u_L\, -\,
\bar{u}_L\gamma_{\mu}s_L\,\bar{d}_L\gamma^{\mu}u_L
\nonumber \\
&&O_2 = \bar{d}_L\gamma_{\mu}s_L\,\bar{u}_L\gamma^{\mu}u_L\, +\,
\bar{u}_L\gamma_{\mu}s_L\,\bar{d}_L\gamma^{\mu}u_L\,+\,
2\,\bar{d}_L\gamma_{\mu}s_L\,\bar{d}_L\gamma^{\mu}d_L+\,
2\,\bar{d}_L\gamma_{\mu}s_L\,\bar{s}_L\gamma^{\mu}s_L\nonumber \\
&&O_3 = \bar{d}_L\gamma_{\mu}s_L\,\bar{u}_L\gamma^{\mu}u_L\, +\,
\bar{u}_L\gamma_{\mu}s_L\,\bar{d}_L\gamma^{\mu}u_L\,+\,
2\,\bar{d}_L\gamma_{\mu}s_L\,\bar{d}_L\gamma^{\mu}d_L-\,
3\,\bar{d}_L\gamma_{\mu}s_L\,\bar{s}_L\gamma^{\mu}s_L\nonumber \\
&&O_4 = \bar{d}_L\gamma_{\mu}s_L\,\bar{u}_L\gamma^{\mu}u_L\, +\,
\bar{u}_L\gamma_{\mu}s_L\,\bar{d}_L\gamma^{\mu}u_L\,-\,
\bar{d}_L\gamma_{\mu}s_L\,\bar{d}_L\gamma^{\mu}d_L\nonumber \\
&&O_5 = \bar{d}_L\gamma_{\mu}\lambda^a s_L(\bar{u}_R\gamma^{\mu}
\lambda^a u_R\, +\, \bar{d}_R\gamma^{\mu}\lambda^a d_R\, +\,
\bar{s}_R\gamma^{\mu}\lambda^a s_R)\nonumber \\
&&O_6 = \bar{d}_L\gamma_{\mu}s_L(\bar{u}_R\gamma^{\mu}u_R\, +\,
\bar{d}_R\gamma^{\mu}d_R\, +\, \bar{s}_R\gamma^{\mu}s_R) \ .
\label{two}
\end{eqnarray}
The Wilson coefficients, $c_i$, are scale-dependent and calculable
perturbatively.  The operators $O_1,\cdots ,O_6$ in Eq. (\ref{two})
have the specific $(flavor, isospin)$ quantum numbers
$(8, 1/2)$, $(8, 1/2)$, $(27, 1/2)$, $(27, 3/2)$, $(8, 1/2)$ and
$(8, 1/2)$, respectively.  The operators $O_{5,6}$, with LR
chiral structure are generated by QCD penguin-type
corrections and, as noted above, have
different chiral structure than do the remaining operators. Of the
operators, $O_{1,\cdots ,6}$, only $O_4$ is $\Delta$I=3/2.  An
important feature of $O_4$ is that it is symmetric in the colors of
the quark fields. Thus, the contributions to a $B^\prime BM$ vertex
(where $M$ is a meson and $B,B^\prime$ are baryons) in which two
quark fields from $O_4$ contract with two quarks from either the
initial or final state baryon vanish in the quark model limit for
baryon structure, in which the quark colors are antisymmetric.  In
this limit, therefore, one would expect the \dith couplings to be
saturated by the so-called ``factorization contribution'' graphs,
in which one quark and one antiquark line from $O_4$ end up in the
meson and the remaining quark and antiquark lines contract
with a quark in the initial and a quark in the final state baryon,
respectively. (This statement is not true for the \dioh operators,
even in the quark model limit, and their matrix elements,
therefore, have not only ``factorization'' contributions, but also
non-factorization or ``internal'' contributions.)  The
factorization contributions, thus, involve baryon-to-baryon and
vacuum-to-meson matrix elements of quark currents and these matrix
elements can, of course, be evaluated either in terms of
experimentally determined baryon form factors and meson decay
constants, or in terms of form factors and decay constants related
to those actually measured by $SU(3)_F$ arguments.

The ``factorization approximation'', which we will employ in what
follows, consists of assuming that, as in the quark model limit,
the \dith amplitudes of a non-leptonic strangeness changing process
are given solely by their factorization contributions. The probable
degree of accuracy of this approximation can be investigated by
comparing ``experimental'' and factorization approximation values
of the known hyperon decay amplitudes.  We have put the word
``experimental'' in quotes here because it is, in fact, a non-
trivial matter to go from the experimental amplitudes to the actual
\dith weak transition amplitudes associated with the operator
$O_4$.  This is because the strong interactions, as a consequence
of the difference of $u$ and $d$ quark masses, themselves violate
isospin, and this, combined with the fact that the basic \dioh
transition amplitude is $\sim 20$ times larger than that for \dith
means that strong interaction mixing effects of only a few \% could
lead to significant enhancements or suppressions in the \dith
amplitudes. An example of such an effect is that associated with
strong-interaction induced $\pi$-$\eta$ and $\Sigma^0$-$\Lambda$
mixing, which drastically alters the \dith amplitudes for
the p-wave $\Lambda\rightarrow$N$\pi$ and $\Xi\rightarrow
\Lambda\pi$ transitions (by $\simeq 400\%$ and $\simeq 100\%$,
respectively\onlinecite{krmi32}).  There would also be, in addition,
strong isospin breaking effects associated with final state
interactions which have not, to the best of our knowledge, been
explored to date.  Allowing only for particle mixing corrections,
one finds that, for a scale of $\sim 1$ GeV, the factorization
approximation for the \dith amplitudes yields\onlinecite{krmi32}:

(1) good fits to the experimental values for the s- and p-wave
$\Xi$ amplitudes and the s-wave $\Sigma$ triangle discrepancy;

(2) an underestimate of the p-wave $\Sigma$ triangle discrepancy by
a factor of 3-4;

(3) an overestimate of the s- and p-wave $\Lambda$ amplitudes by a
factor of 3-4.

\noindent It is important to stress, first, that the experimental errors on
the \dith amplitudes are large (the discrepancies above are
$2\sigma$ at most) and, second, that even if we take these
discrepancies seriously, we do not know whether the shortcoming
represents a problem with the quark model approximation to baryon
structure (i.e. the omission of ``internal'' \dith contributions)
or the effect of yet-to-be-corrected-for strong isospin breaking.
We will, therefore, take the view that the factorization
approximation provides an estimate for the \dith couplings of the
vector mesons which can be expected to be reliable to within what
we hope is a conservatively estimated error of a factor of 3-4.

{}From Eqs.(\ref{QCD}) and (\ref{two}) it
is straightforward to obtain the factorization
contributions to the \dith amplitudes, $A^{(3)}(B\rightarrow
B^\prime V)$, relevant to the process $\Lambda$N$\rightarrow$NN:
\begin{eqnarray}
&&A^{(3)}(\Lambda\rightarrow p\rho^- )={4 \over 3}c_4 K
\langle \rho^-\vert \bar{u}_L\gamma_\mu d_L\vert 0\rangle
\langle p\vert \bar{u}_L\gamma^\mu s_L\vert
\Lambda\rangle\nonumber\\
&&A^{(3)}(p\rightarrow n{K^*}^+ )={4 \over 3}c_4 K
\langle {K^*}^+\vert \bar{s}_L\gamma_\mu d_L\vert 0\rangle
\langle n\vert \bar{d}_L\gamma^\mu u_L\vert
p\rangle\label{couplingsa}
\end{eqnarray}
and
\begin{eqnarray}
&&A^{(3)}(\Lambda\rightarrow n\rho^0 )=-A^{(3)}(\Lambda\rightarrow
p\rho^- )/\sqrt{2}\nonumber\\
&&A^{(3)}(n\rightarrow n{\bar{K^*}}^0 )=-A^{(3)}(p\rightarrow
n{K^*}^+ )
\nonumber\\
&&A^{(3)}(p\rightarrow p{\bar{K^*}}^0 )=A^{(3)}(p\rightarrow n{K^*}^+
)\label{couplingsb}
\end{eqnarray}
where $K=\sqrt{2}G_F\, sin(\theta_C)\, cos(\theta_C)$.

In Eqs. (\ref{couplingsa}) and (\ref{couplingsb}) we have omitted the
weak
$\Lambda$N$\omega$ couplings since the \dith factorization
contributions to these couplings vanish. The weak
$\Lambda$N$\rho$ couplings were obtained
previously\onlinecite{MS95}; the K$^*$ couplings are new.  The
relations in Eq. (\ref{couplingsb}) follow from isospin Clebsch-Gordan
coefficients.  To complete our estimates for the \dith couplings
we, therefore, need only to relate the baryon and meson matrix
elements in Eq. (\ref{couplingsa}) to observable meson decay constants
and baryon form factors.

The meson-to-vacuum matrix elements are straightforward. We have:
\begin{eqnarray}
&&\langle 0\vert V^3_\mu\vert \rho (\epsilon , k)\rangle =
f_\rho m^2_\rho \epsilon^\rho_\mu (k)\nonumber\\
&&\langle 0\vert V^{4-i5}_\mu\vert {K^*}^-(\epsilon ,k)\rangle =
f_{K^*}m^2_{K^*}\epsilon^{K^*}_\mu (k)\label{mdc} \, \ ,
\end{eqnarray}
where $V^a_\mu$ is the usual $SU(3)_F$ octet vector current and
$f_V$ the usual dimensionless vector meson decay constant. {}From
$\rho^0\rightarrow e^+e^-$, $f_\rho =0.2$\onlinecite{PDG}, while
from $\tau^-\rightarrow\nu_\tau {K^*}^-$, $f_{K^*}m^2_{K^*}\simeq
2f_\rho m^2_\rho$.

Similarly, dropping the second class baryon form factors $f_3$ and
$g_2$, and the form factor $g_3$, which yields vanishing
contribution when contracted against the transverse vector meson
polarization vector, one finds that the baryon matrix elements can
be written as:
\begin{equation}
\langle B^\prime (p^\prime )\vert V_\mu -A_\mu\vert B(p)\rangle
\rightarrow \bar{u}_{B^\prime}(p^\prime )\left[ f_1^{BB^\prime }
\gamma_\mu -{\rm i} {\sigma_{\mu\nu}q^\nu\over 2m_N}f_2^{BB^\prime}
+g_1^{BB^\prime}\gamma_\mu\gamma_5\right] u_B(p)\ .\label{bme}
\end{equation}
To obtain values for the baryon form factors not measured
experimentally, we rely on $SU(3)_F$ arguments.  The form factors
$f_1^{BB^\prime}$, $f_2^{BB^\prime}$ are then determined by CVC
(the former in terms of the SU(3) structure constants, the latter
in terms of the neutron and proton anomalous magnetic moments); the
$g_1^{BB^\prime}$ form factors are similarly determined from the
$SU(3)_F$ analysis used for the axial vector hyperon decay
amplitudes, which produces $D$, $F$ factors $D_A =0.79$
and $F_A=0.47$.  (For a clear discussion of this analysis, see, for
example, Ref.\onlinecite{DGHbook}.)

Combining all aspects of the above analysis and defining
the weak \dith $B\rightarrow B^\prime V$ vertex form factors via
\begin{equation}
\langle B^\prime (p^\prime ) V(\epsilon ,k)\vert {\cal
H}_{eff}^{\Delta I=3/2}
\vert B(p)\rangle = \epsilon^{(V)}_\mu
\bar{u}_{B^\prime}(p^\prime )\left[ f_1 \gamma^\mu -{\rm i}
{\sigma^{\mu\nu}q_\nu\over 2m_N}f_2 + g_1\gamma^\mu\gamma_5\right]
u_B(p)\ ,\label{vecamp}\end{equation}
in the factorization approximation, we find the following values for the
weak form factors:
\begin{eqnarray}
&&f_1(\Lambda\rightarrow {\rm p}\rho^-) = -{1\over \sqrt{3}} c_4 K f_\rho
m^2_\rho = -1.2\times
10^{-7}\nonumber\\
&&f_2(\Lambda\rightarrow {\rm p}\rho^-) = 1..63 f_1
(\Lambda\rightarrow {\rm p}\rho^-)
= -1.9\times 10^{-7}\nonumber\\
&&g_1(\Lambda\rightarrow {\rm p}\rho^-) = -0.72 f_1
(\Lambda\rightarrow {\rm p}\rho^-)
= 0.85\times 10^{-7}\label{rhocase}
\end{eqnarray}
for the process
$\Lambda\rightarrow$p$\rho^-$, and:
\begin{eqnarray}
&&f_1({\rm p}\rightarrow{\rm n K}{^*}^{+}) = {1\over 3} c_4 K f_{K^*}
m^2_{K^*}
=
1.4\times
10^{-7}\nonumber\\
&&f_2({\rm p}\rightarrow{\rm n K}{^*}^{+}) = 3.7 f_1
({\rm p}\rightarrow {\rm n K}{^*}^{+})
= 5.1\times 10^{-7}\nonumber\\
&&g_1({\rm p}\rightarrow{\rm n K}{^*}^{+}) = 1.26 f_1
({\rm p}\rightarrow{\rm n K}{^*}^{+})
= 1.7\times 10^{-7}\label{kstarcase}
\end{eqnarray}
for the p$\rightarrow$n K${^*}^{+}$ one,
where, in obtaining the numerical values quoted, we have used
$c_4(1\, {\rm GeV})=0.49$ from Refs.\onlinecite{heffref1,heffref2}.

In light of the discussion above, in exploring the potential consequences of
including \dith contributions to the couplings, we will allow a
variation of up to a factor of 3 about the central values quoted in
Eqs. (\ref{rhocase}) and (\ref{kstarcase}).

The $\Delta$I=3/2 couplings derived above are incorporated in the
hypernuclear weak decay model described in Ref. \cite{PRB97}.
We then apply the modified model to the decays of
${}^{12}_\Lambda$C below.
Enforcing the
$\Delta$I=1/2 rule, Ref. \cite{PRB97} developed a nonrelativistic
$\Delta$S=1
potential for the $\Lambda$N$\to$NN transition in $\Lambda$-hypernuclei
on the basis of a meson-exchange model.
In addition to the long-ranged pion,
the exchange of the other pseudoscalar
mesons belonging to the octet, the $\eta$ and the K,
was included along with
the vector mesons $\rho$, $\omega$ and K$^*$. The
weak coupling constants for the parity-violating (PV) amplitudes were
obtained by making use of soft-meson techniques plus SU(3) symmetry,
which allows one
to relate the unphysical amplitudes of the nonleptonic hyperon
decays involving $\eta$'s and K's to the physical $\mbox{{\rm B} $\to$
{\rm B'} + $\pi$}$
amplitudes. SU(6)$_{\rm W}$, on the other hand, was used to relate the
vector
meson amplitudes to the pseudoscalar meson ones. For the evaluation of
the parity-conserving (PC) amplitudes the pole model was employed.

The nonmesonic decay rate is directly related to the hypernuclear
transition amplitude, in a manner
which facilitates the transition from an initial
hypernuclear state to a final state composed of two nucleons
and a residual (A-2)-particle system. Using standard nuclear structure
methods, the hypernuclear amplitude
can be expressed in terms of two-body amplitudes, $\Lambda$N$\to$NN.
Short range $\Lambda$N correlations were accounted for in
Ref.\cite{PRB97} through an appropriate correlation
function based on the Nijmegen $\Lambda$N interaction\cite{nijme},
while the NN wave function was obtained by solving
the scattering problem of two nucleons moving
under the influence of the strong interaction.

The nonrelativistic reduction of the free space Feynman amplitude
for the virtual meson exchange gives the transition
potential in momentum space.
For vector mesons
the $\Delta$I=1/2 potential reads:

\begin{eqnarray}
V({\bf q})  &=&
G_F m_\pi^2
 \left( F_1 {\hat \alpha} - \frac{({\hat \alpha} + {\hat \beta} )
 ( F_1 + F_2 )} {4M \overline{M}}
(\mbox{\boldmath $\sigma$}_1 \times {\bf q})
(\mbox{\boldmath $\sigma$}_2 \times {\bf q}) \right. \nonumber \\
& & \phantom { G_F m_\pi^2 A }
\left. +i \frac{{\hat \varepsilon} ( F_1 + F_2 )} {2M}
(\mbox{\boldmath $\sigma$}_1 \times
\mbox{\boldmath $\sigma$}_2 ) {\bf q}\right)
\frac{1}{{\bf q}^2 + \mu^2} \
\label{eq:vecpot}
\end{eqnarray}
where $\mu$ is the mass of the meson, $F_1$ and
$F_2$ are the strong vector and tensor couplings,
for which we take those of the Nijmegen soft-core potential \cite{nijme},
and ${\hat \alpha}$, ${\hat
\beta}$ and ${\hat \varepsilon}$ operators which contain, in addition
to the weak coupling constants listed in Table \ref{tab:wcc} (in units
of $G_F m_\pi^2 = 2.21 \times 10^{-7}$) for the
$\rho$ and K$^*$ mesons, the particular
isospin structure.
For the isovector $\rho$-meson the isospin dependence reads:
\begin{eqnarray}
{\hat \alpha}_{\rho} &=& f_1 \,\, \mbox{\boldmath $\tau$}_1
\mbox{\boldmath $\tau$}_2  \nonumber \\
{\hat \beta}_{\rho} &=& f_2 \,\, \mbox{\boldmath $\tau$}_1
\mbox{\boldmath $\tau$}_2  \nonumber \\
{\hat \varepsilon}_{\rho} &=& g_1 \,\, \mbox{\boldmath
$\tau$}_1
\mbox{\boldmath $\tau$}_2 \, {\rm .}
\end{eqnarray}
where now $f_1, f_2$ and $g_1$ (denoted by $\alpha_\rho, \beta_\rho$ and
$\varepsilon_\rho$ in Ref. \cite{PRB97}) correspond to $\Delta$I=1/2
weak $\Lambda$N$\rho$ coupling constants.
{}For the isodoublet K$^*$-meson there
are contributions proportional to $\hat{1}$ and to $\mbox{\boldmath
$\tau$}_1 \mbox{\boldmath $\tau$}_2$ that depend on the
coupling constants, leading to the following dependence:

\begin{eqnarray}
{\hat \alpha}_{K^*} &=&  \frac{
C^{\rm \scriptscriptstyle{PC,V}}_{\rm\scriptscriptstyle{K^*}}} {2} +
D^{\rm \scriptscriptstyle{PC,V}}_{\rm\scriptscriptstyle{K^*}} + \frac{
C^{\rm \scriptscriptstyle{PC,V}}_{\rm\scriptscriptstyle{K^*}}} {2}
\mbox{\boldmath $\tau$}_1
\mbox{\boldmath $\tau$}_2 \nonumber \\
{\hat \beta}_{K^*} &=& \left(
\frac {
C^{\rm \scriptscriptstyle{PC,T}}_{\rm\scriptscriptstyle{K^*}} } {2} +
D^{\rm \scriptscriptstyle{PC,T}}_{\rm\scriptscriptstyle{K^*}} \right) +
\frac{ C^{\rm \scriptscriptstyle{PC,T}}_{\rm\scriptscriptstyle{K^*}}
}{2}
\mbox{\boldmath $\tau$}_1
\mbox{\boldmath $\tau$}_2 \nonumber \\
{\hat \varepsilon}_{K^*} &=&
\frac{
C^{\rm \scriptscriptstyle{PV}}_{\rm\scriptscriptstyle{K^*}}} {2} +
D^{\rm \scriptscriptstyle{PV}}_{\rm\scriptscriptstyle{K^*}} + \frac {
C^{\rm \scriptscriptstyle{PV}}_{\rm\scriptscriptstyle{K^*}}} {2}
\mbox{\boldmath $\tau$}_1
\mbox{\boldmath $\tau$}_2  \ .
\end{eqnarray}

Following  Ref. \cite{PRB97} we also included a monopole form
factor at each vertex.

The $\Delta$I=3/2 amplitudes can be straightforwardly included by assuming
the $\Lambda$ to behave like an isospin $\mid 3/2 \ -1/2 \rangle$
state
and introducing an isospin dependence in the $\Delta$I=3/2
transition potential of the type
$\mbox{\boldmath $\tau$}_{3/2} \mbox{\boldmath $\tau$}$,
where $\mbox{\boldmath $\tau$}_{3/2}$
is the $1/2 \rightarrow 3/2$ isospin
transition operator whose spherical components have the matrix
elements:
\begin{equation}
\langle 3/2\ m^\prime \mid \tau_{3/2}^{(i)} \mid 1/2\ m \rangle =
\langle 1/2\ m\ 1\ i  \mid 3/2\ m^\prime \rangle
\hspace{3cm} i=\pm 1,0 \, \, {\rm .}
\end{equation}

The $\Delta$I=3/2 transition potential for the $\rho$ and K$^*$
mesons is then of the same form as the $\Delta$I=1/2 one but
replacing in Eq. (\ref{eq:vecpot}) the isospin operators by:

\begin{eqnarray}
{\hat \alpha} &\to& f_1 \,\, \mbox{\boldmath $\tau$}_{3/2}
\mbox{\boldmath $\tau$}
\nonumber
\\
{\hat \beta} &\to& f_2 \,\,
\mbox{\boldmath $\tau$}_{3/2}
\mbox{\boldmath $\tau$}
\nonumber
\\
{\hat \varepsilon} &\to& g_1 \,\,
\mbox{\boldmath $\tau$}_{3/2}
\mbox{\boldmath $\tau$} \ .
\label{eq:isop3}
\end{eqnarray}
for both the K$^*$ and the $\rho$, where the weak $\Delta$I=3/2
couplings
$f_1$, $f_2$ and $g_1$, listed in Table \ref{tab:wcc}, are determined by
comparing the values obtained from the operators in Eq. (\ref{eq:isop3})
with those for the specific transitions quoted in Eqs. (\ref{rhocase})
and
(\ref{kstarcase}).

Specifically, for the $\rho$ meson, the operators in Eq.
(\ref{eq:isop3})
give rise to the following product of isospin factors, coming from the
weak and strong vertices:

\begin{eqnarray}
\sqrt{\frac{2}{3}} G_w  \times 1 \,\,\, & &
\mbox{for the } \Lambda {\rm p} \rightarrow {\rm np} \,
\mbox{transition} \nonumber
\\
\frac{1}{\sqrt{3}} G_w \times \sqrt{2} \,\,\, & &
\mbox{for the } \Lambda {\rm p} \rightarrow {\rm pn} \,
\mbox{transition}\nonumber
\\
\sqrt{\frac{2}{3}} G_w \times \left( - 1 \right) \,\,\,& &
\mbox{for the } \Lambda {\rm n} \rightarrow {\rm nn} \,
\mbox{transition} \, \ ,
\label{isosfac}
\end{eqnarray}
where $G_w$ stands for either
$f_1, f_2$ or $g_1$.
Eq. (\ref{isosfac}) demonstrates that, for example, the weak coupling contribution for the
exchange term
$\Lambda$p$\rightarrow$pn is given by $G_w/\sqrt{3}$, indicating that the constants
$f_1, f_2$ and $g_1$ (represented by $G_w$) must be defined as those in
Eq. (\ref{rhocase}), corresponding to the process
$\Lambda \to {\rm p} \rho^- $,
but containing an additional factor $\sqrt{3}$.

For the K$^*$ meson, simple Clebsch-Gordan algebra relations allow us
to see that the transitions $\Lambda {\rm p} \to {\rm np}$ and
$\Lambda {\rm p} \to {\rm pn}$  are identical, while $\Lambda {\rm n}
\to {\rm nn}$ has
a relative change of sign.
This can be easily seen from the relations (\ref{couplingsb}) for the weak
vertex and from the equality of the $\Lambda$NK$^*$ couplings in the
strong sector.
The operator
$\mbox{\boldmath $\tau$}_{3/2} \mbox{\boldmath $\tau$}$
of Eq. (\ref{eq:isop3}),
connecting the initial and final states taking the $\Lambda$ as a
$\mid 3/2\ -1/2 \rangle$ state,
yields precisely this structure introducing an extra factor
$\sqrt{2/3}$.
This global factor will be compensated
by using weak couplings ($f_1, f_2$ and $g_1$) which are those of Eq.
(\ref{kstarcase}) multiplied by $\sqrt{3/2}$.

Our results for ${}^{12}_\Lambda$C
are summarized in Table  \ref{tab:restot}.
The observables include the
nonmesonic decay rate, $\Gamma^{\rm nm}$ (in units of the free
$\Lambda$ decay rate, $\Gamma_\Lambda = 3.8 \times 10^{9} {\rm s}^{-1}$),
the ratio of the neutron to proton induced decay rates,
$\Gamma_{\rm n}/\Gamma_{\rm p}$, which are also given separately,
and
the intrinsic asymmetry parameter, $a_\Lambda$, which is related
via simple angular momentum coupling
to the asymmetry parameter, $A_p$.
As mentioned above, the weak $\Delta$I=3/2 coupling constants
are allowed to vary by up to a factor of $\pm 3$ to account for the limitations
of the factorization model.  Since the relative sign between the $\Delta$I=1/2
and $\Delta$I=3/2 amplitudes is not predicted within the factorization approach
we allow for both possibilities.

As shown in Table \ref{tab:restot}
the total decay rate changes
by at most 6\%
and lies within the error bars of the more recent experimental
result\cite{noumi}.
However, more significant changes are seen for the ratio $\Gamma_{\rm
n}/\Gamma_{\rm p}$
which is enhanced by a factor of two for the combination
$(s_\rho,s_{K^*})=(3,3)$ and decreases by the same amount
for the combination $(s_\rho,s_{K^*})=(-3,-3)$, where $s_{\rho}$ and $s_{K^*}$
denote the scaling factors mentioned above.
This dramatic effect on the neuton- to proton-induced ratio is precisely
of the nature expected once $\Delta$I=3/2 amplitudes are included.
While not affecting the total rate the relative contributions of the
vector mesons in a particular isospin channel
are changed, thus modifying the $\Gamma_{\rm
n}/\Gamma_{\rm p}$ ratio.
{}From Table \ref{tab:restot} it is clear that the largest effect comes from
the $\Delta$I=3/2 amplitudes of the K$^*$ meson.  This comes as no
surprise
since Ref.\cite{PRB97} already found the K$^*$ to be the most important
vector meson contribution.

Analyzing the changes in the neutron- to proton-induced
ratio in more detail we find that its modifications are
due almost entirely to changes in the neutron-induced decay rate,
while the proton-induced channel is barely affected.
Thus, the inclusion of $\Delta$I=3/2 amplitudes does not
alleviate the discrepancy between theory and experiment
 for $\Gamma_{\rm p}$. The measured uncertainties in the $\Gamma_{\rm
n}/\Gamma_{\rm p}$
ratio are too large to draw any conclusions.

While the effects of the $\Delta$I=3/2 amplitudes on the ratio
are significant, their effect on the asymmetry is even more dramatic.
The two extreme values for the asymmetry parameter, which
occur at $(s_\rho,s_{K^*})=(-3,-3)$ and $(s_\rho,s_{K^*})=(3,3)$, differ by
a factor of 7. In contrast to the neutron- to proton-induced ratio,
both the K$^*$ and the $\rho$ contribute about equally to the change
of this observable.  Again, this is consistent with our findings of
Ref.\cite{PRB97} where both the $\rho$ and the $\omega$ significantly
affected the asymmetry.
However, all the values of the hypernuclear
asymmetry at $0^\circ$, ${\cal A}(0^\circ)$,
obtained by multiplying $a_\Lambda$ with the $\Lambda$ polarization
in \hyp{12}{C} and shown in the last column, are within the error
bars of the experimental value. Improved measurements using targets
with higher hypernuclear polarization would be very beneficial to
constrain the factorization approach used here.

We mention in passing that Ref.\cite{oka} has studied $\Delta$I=3/2
contributions to the nonmesonic decay in a direct quark mechanism, induced
by four-point quark vertices in the effective weak Hamiltonian.
As in Ref.~\cite{MS94},
large $\Delta$I=3/2 contributions are found. These help to reproduce
the neutron-induced decay rate. However, the sum of one-pion exchange plus
the direct quark mechanism overpredicts the proton-induced rate.
In a constituent quark picture such as that employed in
Ref.\cite{oka}, moreover, one would expect to have to include the
long-range forces mediated by the exchange of {\it all} pseudoscalar
mesons. As was pointed out long ago by Dubach and collaborators
\cite{dubach,dubach96}, and recently confirmed in a rigorous finite-nucleus
calculation \cite{PRB97},
there is considerable cancellation between one-pion-exchange and
one-kaon-exchange contributions
(the latter appearing through diagrams
involving the weak NNK vertex, which contribution is omitted in
Ref.\cite{oka}).
This cancellation
was also seen to be important in the
results of Ref.\cite{MS94}, which included both one-pion-exchange
and one-kaon-exchange, in addition to the direct quark contributions
also present in Ref.\cite{oka}.

In conclusion,
we have studied the effect of the $\Delta$I=3/2 amplitudes in
the meson-exchange model of Ref. \cite{PRB97}.
Our results clearly indicate that both the
neutron-induced rate and the asymmetry are
very sensitive
to the presence of $\Delta$I=3/2 amplitudes. Thus, better data for these
observables will help in assessing the validity of the $\Delta$I=1/2 rule
for the weak vector meson sector.  Hopefully, such data will
be forthcoming soon from the FINUDA facility at DAPHNE.

AP and AR acknowledge Joan Soto for clarifying discussions regarding
the effective weak Hamiltonian.
The work of CB was supported by US-DOE grant no. DE-FG02-95-ER40907 while
the work of AP and AR was supported by
DGICYT contract no. PB95-1249 (Spain) and by the Generalitat de
Catalunya grant no. GRQ94-1022.
This work has received support from the NATO Grant
CRG 960132.
AP acknowledges support
from a doctoral fellowship of the Ministerio de Educaci\'on y Ciencia
(Spain). KM acknowledges the ongoing support of the Natural
Sciences and Engineering Research Council of Canada, and the
hospitality of the Special Research Center for the Subatomic
Structure of Matter of the University of Adelaide.

\begin{table}
\centering
\caption{Weak coupling constants for the $\rho$ and K$^*$
exchange potential in units of $G_F m_\pi^2 = 2.21 \times
10^{-7}$}
\vskip 0.1 in
\begin{tabular}{lrrrr}
   & \multicolumn{2}{c}{$\rho$} & \multicolumn{2}{c}{K$^*$}
\\
   & $\Delta$I=3/2 & $\Delta$I=1/2 & $\Delta$I=3/2 & $\Delta$I=1/2 \\
\hline
$f_1$ & $-0.93$ & $-3.50$ & 0.77 & $C^{\rm PC,V}_{{\rm K}^*}$ =
$-3.61$ \\
  &  &  &  & $D^{\rm PC,V}_{{\rm K}^*}$ = $-4.89$ \\
\hline
$f_2$ & $-1.52$ & $-6.11$ & 2.83 & $C^{\rm PC,T}_{{\rm K}^*}$ =
$-17.9$ \\
  &  &  &  & $D^{\rm PC,T}_{{\rm K}^*}$ = $9.30$ \\
\hline
$g_1$ & 0.67 & 1.09 & 0.94 & $C^{\rm PV}_{{\rm K}^*}$ =
$-4.48$ \\
  &  &  &  & $D^{\rm PV}_{{\rm K}^*}$ = 0.60 \\
\end{tabular}
\label{tab:wcc}
\end{table}

\begin{table}
\centering
\caption{Weak decay observables of
${}^{12}_\Lambda$C for
the full meson exchange potential
 }
\vskip 0.1 in
\begin{tabular}{rrccccrr}
$s_\rho$ & $s_{K^*}$ & $\Gamma^{\rm nm}/\Gamma_\Lambda$ &
$\Gamma_{\rm n}/\Gamma_{\rm p}$ &
$\Gamma_{\rm n}/\Gamma_\Lambda$ &
$\Gamma_{\rm p}/\Gamma_\Lambda$ &
$a_\Lambda$ & ${\cal A}(0^\circ)$
\\
\hline
$-3$ & $-3$ & 0.783 & 0.034 & 0.026 & 0.758 & $-0.522$ & $-0.050$ \\
$-3$ & 0 & 0.757 & 0.061 & 0.043 & 0.714 & $-0.414$ & $-0.039$ \\
$-3$ & 3 & 0.789 & 0.115 & 0.082 & 0.708 & $-0.280$ & $-0.030$ \\
0 & $-3$ & 0.779 & 0.040 & 0.030 & 0.748 & $-0.434$ & $-0.041$ \\
0 & 0 & 0.753 & 0.068 & 0.048 & 0.705 & $-0.316$ & $-0.030$ \\
0 & 3 & 0.786 & 0.123 & 0.086 & 0.700 & $-0.178$ & $-0.017$ \\
3 & $-3$ & 0.787 & 0.053 & 0.039 & 0.747 & $-0.338$ & $-0.032$ \\
3 & 0 & 0.762 & 0.081 & 0.057 & 0.705 & $-0.212$ & $-0.020$ \\
3 & 3 & 0.796 & 0.136 & 0.095 & 0.701 & $-0.072$ & $-0.007$ \\
\hline
EXP: & & 1.14$\pm$0.2\cite{szymanski} & 1.33$^{+1.12}_{-0.81}$
\cite{szymanski} &  & 0.31$^{+0.18}_{-0.11}$\cite{noumi} & &
$-0.01\pm0.10$\cite{ajim} \\
 & & 0.89$\pm$0.15$\pm$0.03\cite{noumi} &
1.87$\pm$0.59$^{+0.32}_{-1.00}$\cite{noumi} & & & & \\
 & &  & 0.70$\pm$0.3\cite{montwill} & & & & \\
 & &  & 0.52$\pm$0.16\cite{montwill} & & & & \\
\end{tabular}
\label{tab:restot}
\end{table}

\end{document}